# PLANETARY-MASS DARK MATTER?

**Abstract:** Sahu et al. recently reported tentative evidence for a large population of "free-floating" planetary-mass objects in a nearby globular cluster. If verified, this discovery is consistent with a prediction that the galactic dark matter has a substantial planetary-mass component.

In 1987 an unorthodox fractal paradigm predicted that gravitational microlensing studies would reveal a large population of isolated planetary-mass lenses that are an important component of the galactic dark matter.[1] A recent report in *Nature* by Sahu *et al.*[2] of tentative evidence for a large population of "free-floating" planetary-mass lenses in one of the Galaxy's globular clusters may be relevant to this prediction.

Sahu *et al.* have demonstrated the advantages of using globular clusters and the Hubble Space Telescope for a second generation of more accurate microlensing experiments. They report one convincing stellar-mass event with an estimated mass of $\approx 0.13$ $M_\odot$. Six other candidates with unresolved light curves (< 1 day) imply a large population of isolated planetary-mass lenses. If these events are due to microlensing, then the lenses have upper limit masses of $\approx 0.25$ $M_{Jup}$.

The Self-Similar Cosmological Paradigm (SSCP) predicts[1] that the mass spectrum of the galactic dark matter (DM) is dominated by two peaks: one at $\approx 0.15$ $M_\odot$ and the other at $\approx 0.05$ $M_{Jup}$. These objects are predicted to be in ultracompact states, i.e., "primordial black holes." For a sufficiently representative sample, both peaks are expected to have roughly equal *numbers* of objects, but most of the dark matter *mass* is in the stellar-mass objects. Previous microlensing experiments have yielded consistent evidence for stellar-mass ($0.1 M_\odot$ to $0.5 M_\odot$) DM objects,[3-5] but the evidence for planetary-mass DM objects has been less convincing.[5,6] The new evidence for planetary-mass lenses presented by Sahu *et al.* is preliminary. Their experiment needs to be repeated with a finer time resolution in order to clarify the true nature of these short-duration events. If future microlensing studies verify the existence of a large class of isolated planetary-mass lenses with a sharp mass peak at $\approx 0.05\ M_{Jup}$, then the status of the SSCP may need to be reevaluated.

Robert L. Oldershaw

Amherst College
Amherst, MA 01002
USA
rlolders@unix.amherst.edu

**Summer correspondence address:**

12 Emily Lane
Amherst, MA 01002
USA